\documentclass[prl,showpacs,preprint,superscriptaddress]{revtex4}
\usepackage{amssymb,graphicx,amsbsy}
\usepackage{bm}

\begin{document}

\title{True and quasi long-range order in the generalized
$q$-state clock model}
\author{Seung Ki Baek}
\affiliation{Department of Physics, Ume{\aa} University, 901 87 Ume{\aa},
Sweden}
\author{Petter Minnhagen}
\affiliation{Department of Physics, Ume{\aa} University, 901 87 Ume{\aa},
Sweden}
\author{Beom Jun Kim}
\email[Corresponding author, E-mail: ]{beomjun@skku.edu}
\affiliation{BK21 Physics Research Division and Department of Energy Science,
Sungkyunkwan University, Suwon 440-746, Korea}

\begin{abstract}
From consideration of the order-parameter distribution,
we propose an observable which makes a clear distinction between
true and quasi long-range orders in the two-dimensional
generalized $q$-state clock model.
Measuring this quantity by Monte Carlo simulations for $q=8$, we
construct a phase diagram and identify critical properties across
the phase-separation lines among the true long-range order, quasi
long-range order, and disorder. Our result supports the theoretical
prediction that there appears a discontinuous order-disorder transition as
soon as the two phase-separation lines merge.
\end{abstract}


\maketitle

The existence of quasi long-range order (LRO) characterizes the
critical behavior of the two-dimensional $XY$ model~\cite{kos,kt,petter1}
as well as its dual, the solid-on-solid (SOS) model to describe the
roughning transition on a surface~\cite{knops,hasen-b,hasen-sos1,hasen-sos2}.
By quasi LRO, we mean that the spin-spin correlation function decays
algebraically, which implies that the system is not magnetically ordered.
We will refer to a phase having such characteristics as
quasiliquid~\cite{lapilli}.
The lack of true magnetic order for the $XY$ model is attributed to
spin-wave excitations,
which are gapless and thus excited at any finite temperature.
On the other hand, the quasi LRO is broken by the vortex-pair unbinding at
the Kosterlitz-Thouless (KT) transition which exhibits an essential
singularity.
Even though the $XY$ model assumes the continuous U(1) symmetry in
the spin angle $\theta$, essentially the same nature is observed when the angle
is discretized into $q$ possible values over $\theta = 0, \frac{2\pi}{q},
\ldots, \frac{2\pi (q-1)}{q}$, as long as $q$ is high enough.
Such a discrete-spin system is called the $q$-state clock model
if two neighboring spins, which have $\theta_i = 2\pi n_i / q$ and
$\theta_j = 2 \pi n_j / q$ with integers $n_i$ and $n_j$, respectively,
interact via cosine potential $V(\theta_i-\theta_j) = -J \cos (\theta_i -
\theta_j)$ with a ferromagnetic coupling constant $J>0$. One can
generalize this interaction with preserved symmetry, $V(\theta) =
V(-\theta) = V(\theta + 2\pi)$, into the form given by the Hamiltonian
\begin{equation}
H = \sum_{\left< i, j \right>} V_p(\theta_i - \theta_j)
= \sum_{\left< i, j \right>} \frac{2J}{p^2} \left[ 1-\cos^{2p^2} \left(
\frac{\theta_i - \theta_j}{2} \right) \right],
\label{eq:h}
\end{equation}
where $\theta_i$ is the $i$th spin angle, and the sum runs over
nearest neighbors~\cite{domany}.
It recovers the $q$-state clock
model at $p=1$ and approaches the $q$-state Potts model in the limit of
large $p$~\cite{wu}.
We denote the system defined by
Eq.~(\ref{eq:h}) as the generalized $q$-state clock model.
Since it has been claimed that this model with $p=1$ and $q \ge 8$ precisely
reproduces the KT transition~\cite{lapilli}, we set $q=8$ throughout this
work. At the same time, the
discreteness introduces a finite gap in the spin-wave excitation, making
the true LRO realizable at low temperatures~\cite{jkkn,elit,cardy,domany1}.
These two phase transitions are connected by the duality relation, which is
exactly established within the Villain approximation~\cite{savit}.
While the appearance of the quasi LRO is readily detected by observables such as
Binder's fourth-order cumulant~\cite{binder,hasen-b} or helicity
modulus~\cite{nelson,petter2,petter3} that of the true LRO has been
observed by changes in specific heat or
magnetization~\cite{domany1,challa1,lapilli}.
It is, however, rather hard to locate the transition temperature
using these quantities, especially for high $p$ values where the quasi
LRO exists in a very narrow temperature range.
Thus alternative quantities are required, for example, like a 
direct observation of the formation of giant
clusters~\cite{tomita}.
In this Rapid Communication, we show that the transition can be well localized by a non-local
order parameter which is obtained from the average spin direction
and which makes a clear distinction between the true and quasi LROs.
Using this quantity it is shown that the quasiliquid phase
disappears beyond $p \simeq 2.8$, where the transition becomes
discontinuous just as for the eight-state Potts model.

Let us consider the generalized eight-state clock model given above on the
$L \times L$ square lattice with the system size $N=L^2$.
The complex order parameter of this
system is defined as
\begin{equation}
m = N^{-1} \sum_j e^{i \theta_j} = |m| e^{i \phi}.
\label{eq:m}
\end{equation}
\begin{figure}
\includegraphics[width=0.3\textwidth]{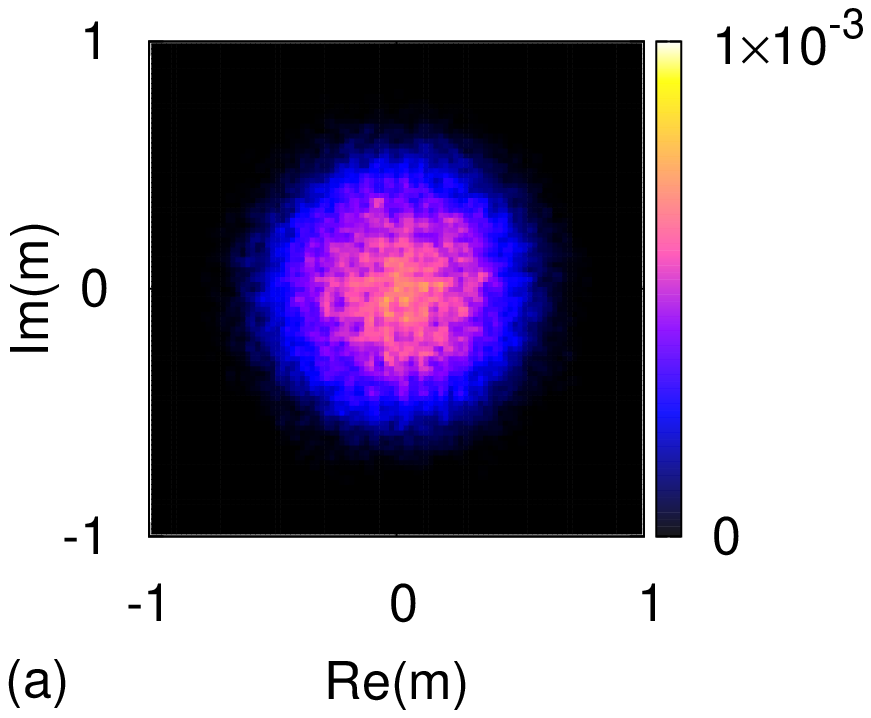}
\includegraphics[width=0.3\textwidth]{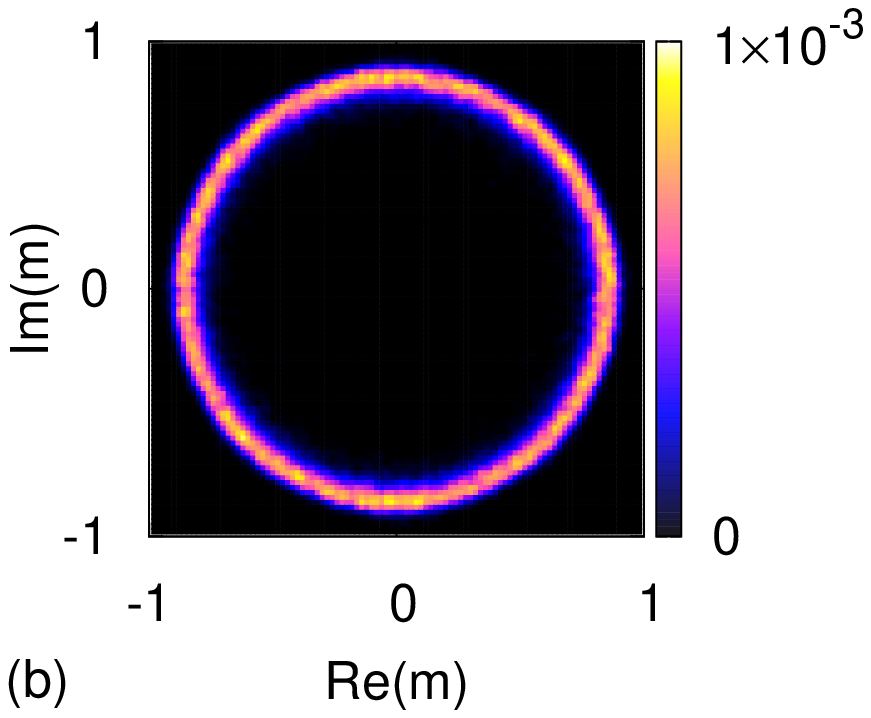}
\includegraphics[width=0.3\textwidth]{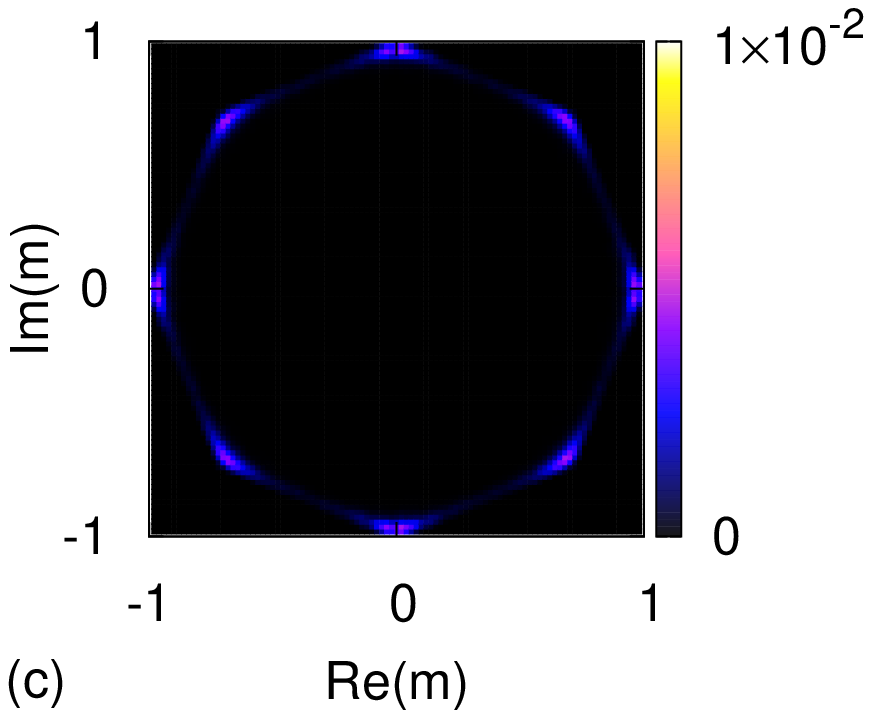}
\caption{(Color online) Distributions of the order parameter $m$ on the complex
plane, obtained for $p=1$, $q=8$, and $L=8$. We start from the
high-temperature regime and then cool down the system slowly.
Here are shown three characteristic distributions, where one finds
(a) a disordered phase at temperature $T=1.50$,
(b) a quasiliquid phase at $T=0.70$, and
(c) an ordered phase at $T=0.36$,
where the temperatures are given in units of $J/k_B$.}
\label{fig:nova}
\end{figure}
As in Ref.~\cite{katz}, it is instructive to visualize the distribution of
$m$ on the complex plane. The distributions in Fig.~\ref{fig:nova} are
obtained by running Monte Carlo simulations with the single-cluster update
algorithm~\cite{wolff,janke,hasen-sto}, and each panel represents a
different phase of the eight-state
clock model at a different temperature.
In the leftmost panel [Fig.~\ref{fig:nova}(a)],
we see the disordered phase in the high-temperature regime.
The order parameter $m$ exhibits a two-dimensional
Gaussian peak around the origin, which may be regarded as a
delta peak at $|m|=0$ in the thermodynamic limit.
Figure~\ref{fig:nova}(b) illustrates the quasiliquid phase,
where the order parameter rotates in the $\phi$ direction with nonzero
magnitude. Note that both of the distributions in Figs.~\ref{fig:nova}(a) and
\ref{fig:nova}(b) manifest a continuous rotational symmetry,
which is spontaneously broken at a lower
temperature as shown in Fig.~\ref{fig:nova}(c). One finds a true LRO being
established so that $m$ indicates well-defined directions selected from
the eight-fold symmetry.

\begin{figure}
\includegraphics[width=0.48\textwidth]{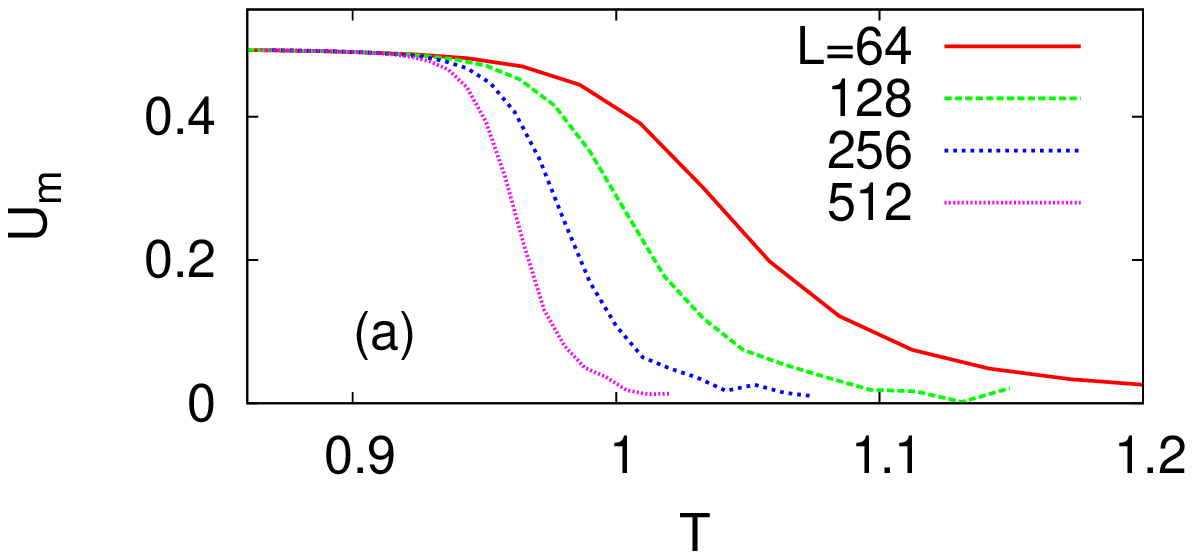}
\includegraphics[width=0.48\textwidth]{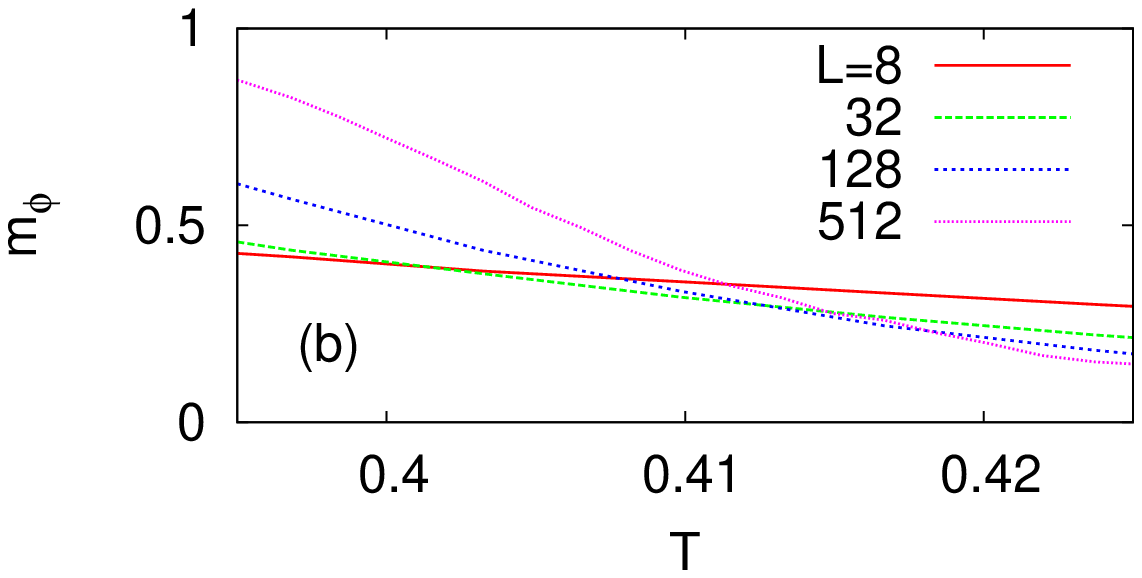}
\includegraphics[width=0.48\textwidth]{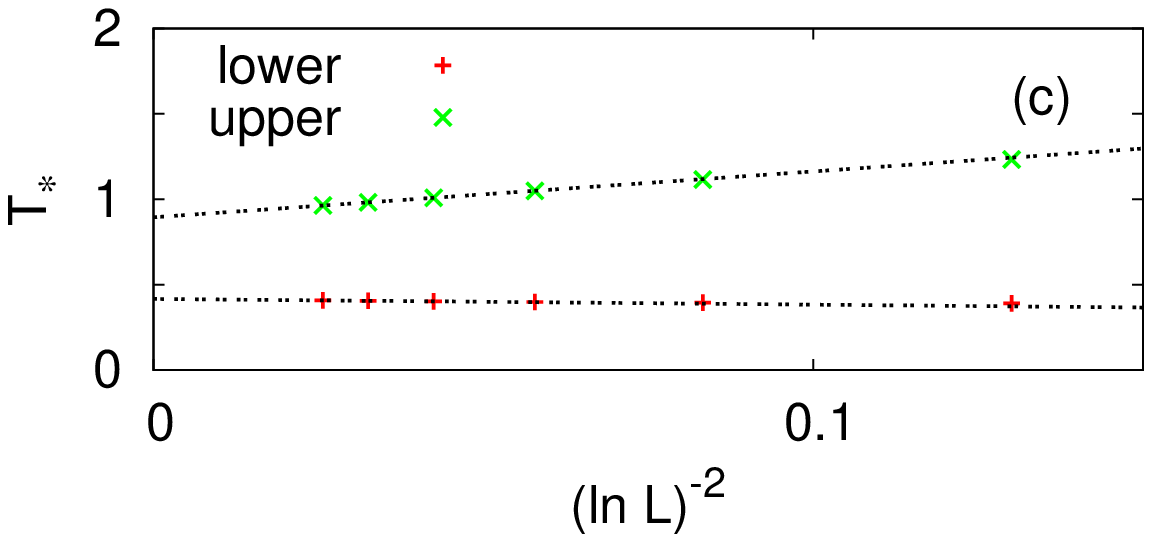}
\includegraphics[width=0.48\textwidth]{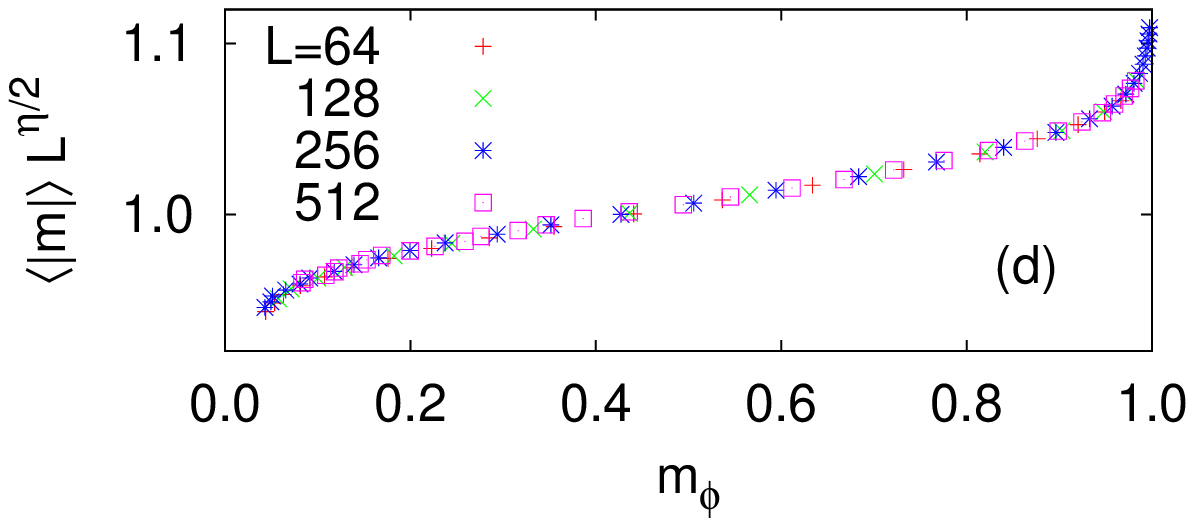}
\caption{(Color online) Double phase transitions for $p=1$ and $q=8$.
(a) The transition between quasiliquid and disordered phases is detected by
merging of $U_m$ curves. (b) The other transition occurs between the ordered and
quasiliquid phases, which is detected by $m_{\phi}$. (c) Extrapolating
positions of inflection points according to the KT picture, we get $T_{c1} =
0.417(3)$ and $T_{c2} = 0.894(1)$. (d) Checking Eq.~(\ref{eq:mm}) with
$\eta/2 = 0.031$, which best describes the data with a single curve.
}
\label{fig:q8}
\end{figure}

A major difference between
Figs.~\ref{fig:nova}(a) and \ref{fig:nova}(b)
lies in the distributions of $|m|$.
The transition between the quasiliquid and disordered phases can be
detected by means of Binder's fourth-order cumulant,
\begin{equation}
U_m = 1 - \frac{\left< |m|^4 \right>}{2 \left< |m|^2 \right>^2},
\label{eq:cum}
\end{equation}
where $\left< \cdots \right>$ represents the thermal average
[Fig.~\ref{fig:q8}(a)].
The factor of two in the denominator of Eq.~(\ref{eq:cum}) is based on
the fact that $\left< |m|^4 \right> = 2\left< |m|^2 \right>^2$ for
such a two-dimensional Gaussian distribution as in Fig.~\ref{fig:nova}(a).
We should note that $U_m$ does not detect the transition between the ordered
and quasiliquid phases since they differ only in the {\em angular}
direction on the complex plane.
Henceforth, we need a quantity capturing the change
along $\phi$. In the same spirit as $U_m$, one may define
a cumulant as
\begin{equation}
U_\phi = 1-\frac{5\left< \tilde\phi^4 \right>}{9\left< \tilde\phi^2
\right>^2},
\label{eq:uphi}
\end{equation}
where $\tilde\phi \equiv (2\pi)^{-1}(q\phi {\rm ~mod~} 2\pi)$ so that
$U_{\phi}$ goes to zero when the distribution is uniform with respect to
$\phi$. Or we may alternatively have
\begin{equation}
m_{\phi} = \left<\cos (q \phi)\right>,
\label{eq:ang}
\end{equation}
which yields a finite value when $\phi$ is
frozen but again vanishes when $\phi$ is
isotropically distributed [Fig.~\ref{fig:q8}(b)].
Provided that the system is nearly ordered with large enough $q$, we
approximately have $\phi \approx N^{-1} \sum_j \theta_j$
from Eq.~(\ref{eq:m}) so that
$m_{\phi} \approx \left< \cos (2\pi \bar{n}) \right>$ with $\bar{n} = N^{-1}
\sum_j n_j$. By duality, the integer field $n_j$ can be mapped to a charge
distribution in the lattice Coulomb gas~\cite{savit}
and the approximate expression for $m_{\phi}$ has been introduced in
Ref.~\cite{hasen-sos2} to monitor the fugacity of charged particles under
numerical renormalization-group calculations.
Since the quasiliquid phase exists between the ordered and
disordered phases for $q=8$, we have two separate transitions at $T=T_{c1}$
and $T_{c2}$, which are clearly detected by the above quantities.
Note the movements of data points in Figs.~\ref{fig:q8}(a) and
\ref{fig:q8}(b) with different system sizes.
Since the position of an inflection point, $T_\ast$, would correspond to
where the transition occurs in the thermodynamic limit,
we may extrapolate them according to the KT scenario,
\begin{equation}
\ln L \sim \left| T_\ast - T_c \right|^{-1/2},
\label{eq:ktl}
\end{equation}
to estimate the critical temperatures both for
the upper and lower transitions [Fig.~\ref{fig:q8}(c)].
In addition, regarding $\left< |m|
\right> = L^{-\eta/2} \tilde{m}(L,T)$ around $T_{c1}$~\cite{challa},
we replace the dependency on both of $L$ and $T$ by that on a single
variable $m_{\phi} = m_{\phi}(L,T)$ so that
\begin{equation}
\left< |m| \right> = L^{-\eta/2} f(m_{\phi}).
\label{eq:mm}
\end{equation}
In other words, plotting $\left< |m| \right> L^{\eta/2}$ against $m_{\phi}$,
data from different sizes are expected to fall on a single curve if one
correctly selects $\eta$. This provides a way to determine $\eta$ even
without precise knowledge of $T_{c1}$ (see also Ref.~\cite{loison}). The
best fit is found at $\eta/2 = 0.031(2)$ as shown in Fig.~\ref{fig:q8}(d),
while the theoretical value is given as $\eta/2 = 1/32 = 0.03125$ at
$T=T_{c1}$~\cite{elit}.

\begin{figure}
\includegraphics[width=0.48\textwidth]{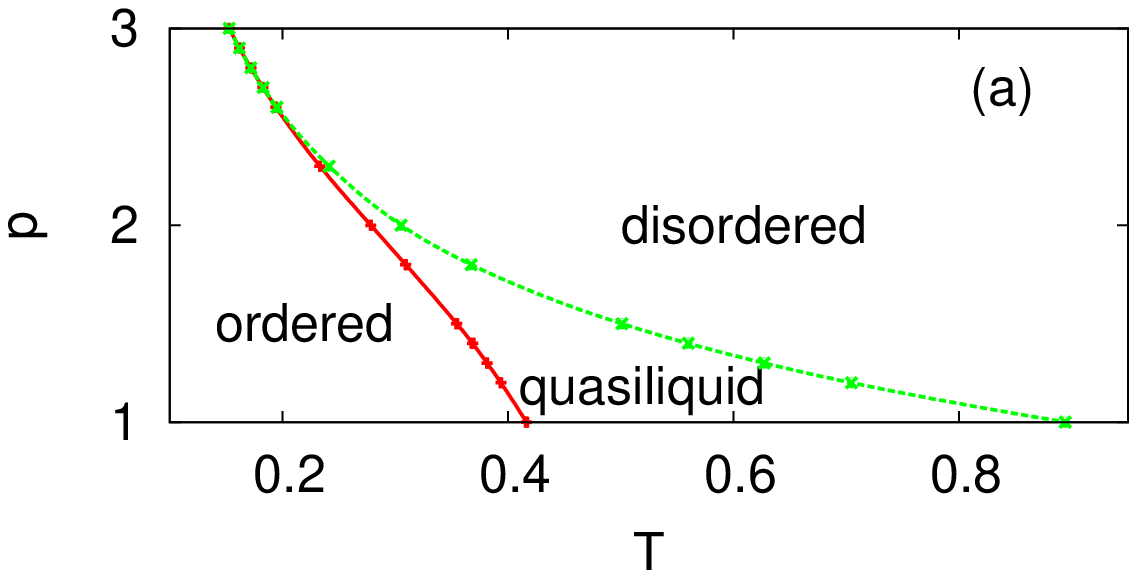}
\includegraphics[width=0.48\textwidth]{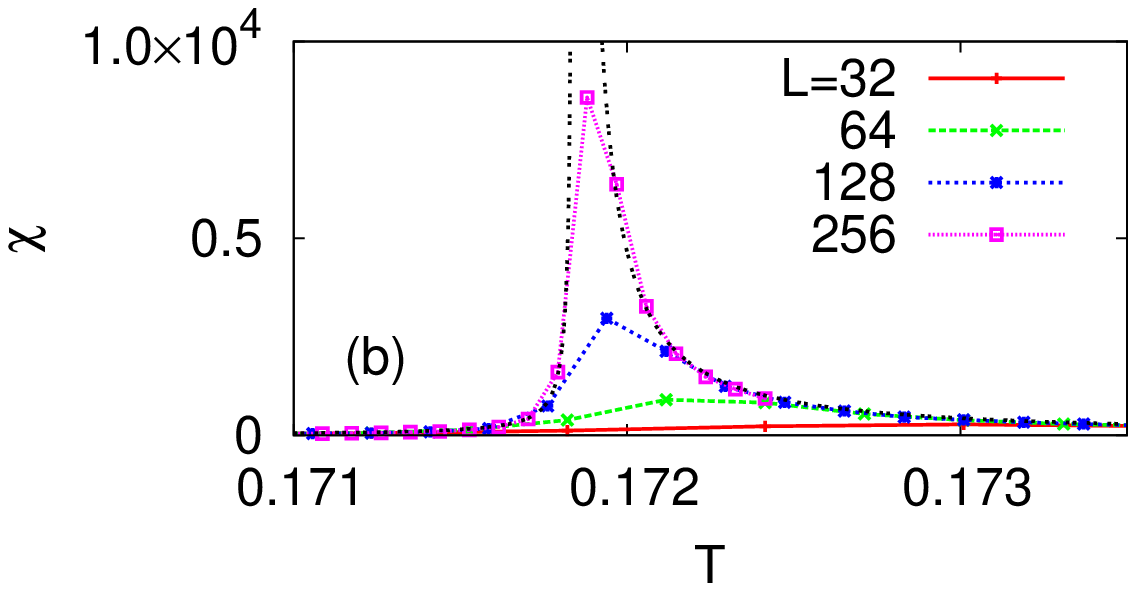}
\includegraphics[width=0.48\textwidth]{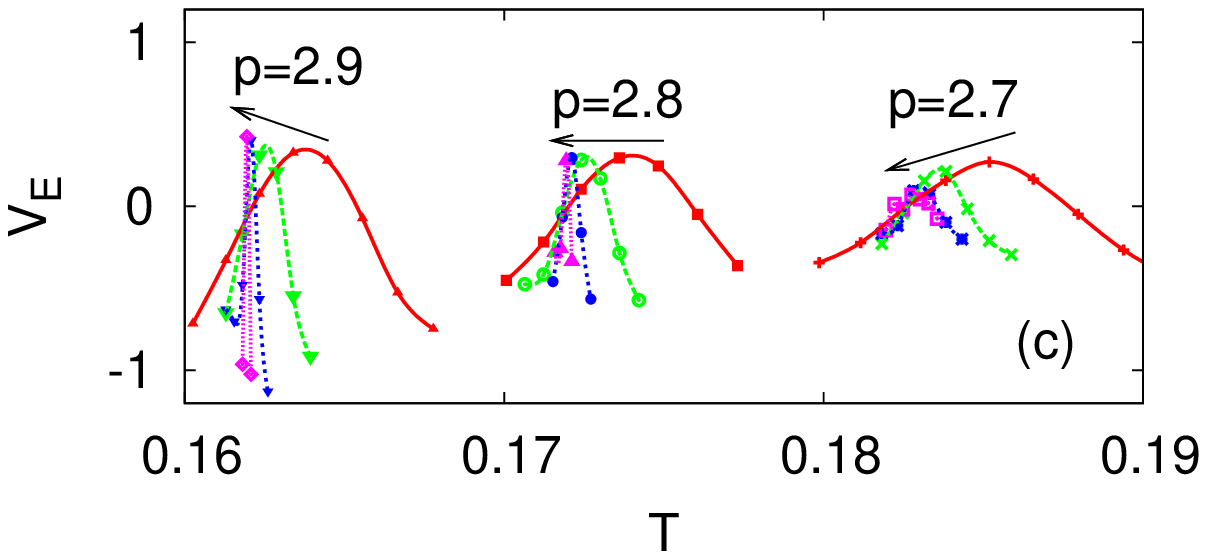}
\includegraphics[width=0.48\textwidth]{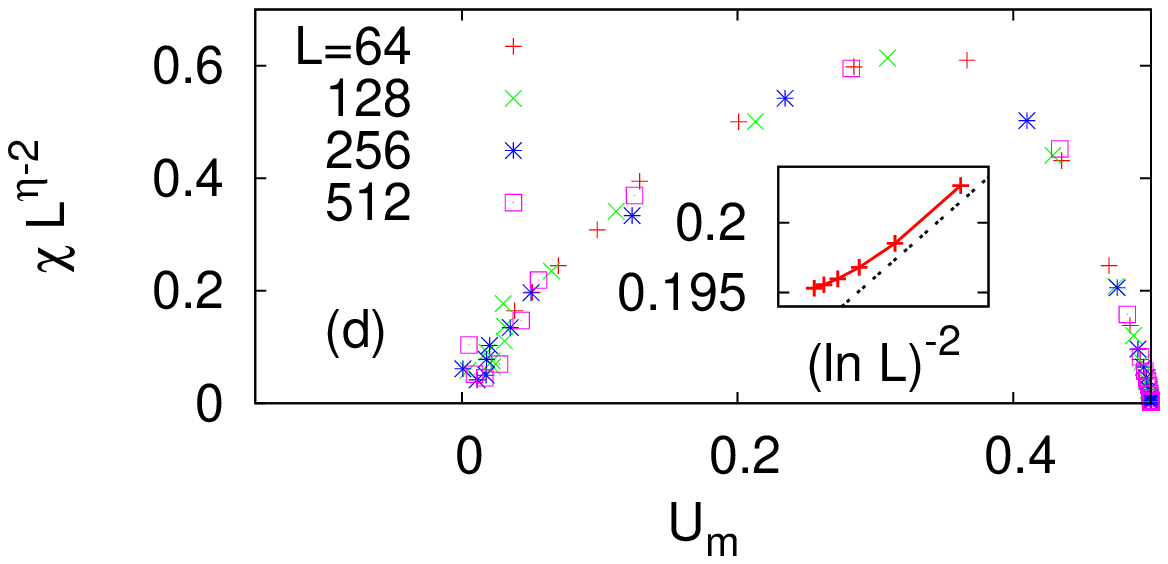}
\caption{(Color online) (a) Phase diagram of the generalized eight-state
clock model. (b) Susceptibility as a function of $T$ at $p = 2.8$, where
the dotted lines describe $\chi \sim |T-T_c|^{-1.2}$, with $T_c = 0.17184$.
(c) $V_E$ as a function of $T$. For each $p$, system sizes are given by
$L=16, 32, 64$, and $128$ from right to left.
(d) Estimation of $\eta$ for $p=2.6$ by Eq.~(\ref{eq:xu}). Seventh-order
polynomials are used to find the best fit, from which $\eta=0.41(4)$ is
estimated. Inset: $T_{c2}$ at $p=2.6$ against $(\ln L)^{-2}$ up to $L=512$.
It moves slower than predicted by Eq.~(\ref{eq:ktl})
as shown in comparison with the straight dotted line.
}
\label{fig:ph}
\end{figure}

When altering the potential shape by increasing $p$ in Eq.~(\ref{eq:h}),
one may well expect that the two transitions will eventually transform into a
single, discontinuous transition at a certain $p$ value as the Potts-model
limit is approached. How this happens can be found by
numerical simulations, and a phase diagram thereby obtained is shown
in Fig.~\ref{fig:ph}(a). 
It seems that the two phase-separation lines merge as $p$ approaches $3.0$.
A better estimate is obtained by looking at magnetic susceptibility.
Recalling that susceptibility $\chi = N (k_B T)^{-1} \left( \left< |m|^2
\right> - \left< |m| \right>^2\right)$ corresponds to the sum of correlations,
we may argue that its divergence implies long-ranged correlations over the
system, a key feature of the quasiliquid phase.
If $p$ is small enough to exhibit the quasiliquid phase, susceptibility
indeed diverges over a finite temperature range. For $p=2.8$, however,
we find that data points fall on $\chi \sim |T-T_c|^{-1.2}$ which has only
one singular point at $T=T_c$ [Fig.~\ref{fig:ph}(b)]. This
$T_c$ is also consistent with the results from $U_m$ and $m_{\phi}$.
We therefore conclude that the quasiliquid phase shrinks
to a single point at $p \simeq 2.8$. Furthermore, the distribution of energy
per spin, $E$, exhibits double peaks for $p \gtrsim 2.8$.
By analogy with $U_m$, we introduce the following quantity:
\begin{equation}
V_E = 1 - \frac{\left< \left(E-\left<E\right>\right)^4 \right>}{3\left<
\left(E - \left<E\right> \right)^2 \right>^2}.
\label{eq:ve}
\end{equation}
Recall that if a scalar variable $x$ has a one-dimensional
Gaussian distribution with zero mean, one readily finds
$\left< x^4 \right> = 3 \left< x^2 \right>^2$. Consequently, $V_E$ will
vanish when there exists a single peak positioned at $\left< E \right>$.
It will approach a nontrivial value, however, when the energy
distribution has double peaks on opposite sides of the average value $\left<
E \right>$. A similar attempt to define such a quantity has already been
made in Ref.~\cite{challa} for characterizing a discontinuous transition.
Figure~\ref{fig:ph}(c) shows that $V_E$ remains finite at $p \gtrsim 2.8$,
which signals a change to a discontinuous transition~\cite{cardy,domany1}.

The concept of universality suggests that the critical properties will be kept
the same in the vicinity of $p=1$. However, one may ask if the natures of the
transitions between the ordered and quasiliquid phases and between the
quasiliquid and disordered phases depend on the value of $p$.
Applying Eq.~(\ref{eq:mm}) to higher $p$ values, we find that $\eta = 1/16$
cannot be ruled out even when $p$ approaches $2$. However, the quality
of fit severely deteriorates at $p$ higher than $2$, possibly due to that our
magnetization data are easily influenced by the proximity of the upper
transition.
On the other hand, with the same motivation as in Eq.~(\ref{eq:mm}),
we may characterize the upper transition
by means of the following scaling relation~\cite{loison}:
\begin{equation}
\chi = L^{2-\eta} g(U_m).
\label{eq:xu}
\end{equation}
This method yields $\eta = 0.24(1)$ at $p=1.0$ in agreement with
the prediction of $1/4 = 0.25$ for the KT transition~\cite{kos}.
It is not very surprising that $\eta$ tends to be
underestimated here if taking into account the logarithmic
correction involved in susceptibility~\cite{kenna,hasen}. We
observe from our numerical data that the criticality deviates from
the standard KT type below the merging point. If we take $p=2.6$, for
instance, the best fit is found at $\eta = 0.41$ and the
size dependence of the transition temperatures deviates from
Eq.~(\ref{eq:ktl}) [Fig.~\ref{fig:ph}(d)]. Still, it remains to be
investigated in detail how the critical behavior begins to change
or if the standard KT behavior is recovered for even larger lattice sizes
($L>512$) in spite of the data collapse shown in Fig.~\ref{fig:ph}(d) for
lattice sizes up to $L=512$.

In summary, we have proposed a practical quantity to distinguish the true
and quasi LROs, based on the order-parameter distribution. Using this
quantity, we have provided a phase diagram on the $p-T$ plane for the
generalized eight-state clock model.
It has been shown that a
discontinuous transition appears when the phase-separation lines merge into
one at $p \simeq 2.8$. We have also checked critical properties along the
lines and found changes in scaling behaviors before reaching
the merging point from numerical calculations up to $L=512$.

\acknowledgments
S.K.B. and P.M. acknowledge the support from the Swedish Research Council
with the Grant No. 621-2002-4135, and
B.J.K.  was supported by WCU(World Class University) program through the
National Research Foundation of Korea funded by the Ministry of Education,
Science and Technology (Grant No. R31-2008-000-10029-0).
This work was conducted using the resources of High Performance
Computing Center North (HPC2N).


\end{document}